\documentclass[seceq,twocolumn]{jpsj2} 
\title{
\begin{center}
{\large {\bf Quadrupole Susceptibility and Elastic Softening \\due to a Vacancy in Silicon Crystal } }
\end{center}}

\author{
Takemi {\sc Yamada}\thanks{E-mail address: takemi@phys.sc.niigata-u.ac.jp}, 
Youichi {\sc Yamakawa} and Yoshiaki {\sc \=Ono}
}

\inst{Department of Physics, Niigata University, Ikarashi, Nishi-ku , Niigata, 950-2181, Japan}

\abst{
We investigate the electronic states around a single vacancy in silicon
crystal by using the Green's function approach. The triply degenerate
vacancy states within the band gap are found to be extended over a large
distance $\sim20$ {\AA} from the vacancy site and contribute to the
reciprocal temperature dependence of the quadrupole susceptibility
resulting in the elastic softening at low temperture. The Curie constant
of the quadrupole susceptibility for the trigonal mode
($O_{yz},O_{zx},O_{xy}$) is largely enhanced as compared to that for the
tetragonal mode ($O_{2}^{0},O_{2}^{2}$). The obtained results are
consistent with the recent ultrasonic experiments in silicon crystal
down to 20 mK. We also calculate the dipole and octupole
susceptibilities and find that the octupole susceptibilities are
extremely enhannced for a specific mode.
}

\kword{silicon, vacancy, softening, quadrupole, elastic constant}

\begin{document}
\maketitle
\section{Introduction}
The recent discovery of low temperature elastic softening in silicon (Si)
crystal\cite{ob-1} has stimulated much interest in the investigations
not only for physical properties but also for industrial applications as
the softening is closely related to the Si vacancy
concentration\cite{ob-2}. The elastic constants of the both
$(C_{11}-C_{12})/2$ and $C_{44}$ modes show reciprocal temperature
dependence below 20 K down to 20 mK, which implies the existence of
triply degenerate groundstates which couple to the strain caused by the
ultrasound\cite{ob-1}. The softening of non-doped Si is independent
of the external magnetic fields up to 10 T and is attributed to the
vacancy with the non-magnetic neutral charge state $V^0$, while that of
boron (B)-doped Si is suppressed by the external magnetic fields of
the order of 1 T and is attributed to the vacancy state $V^+$ with the
valence $+1$ and the spin 1/2\cite{ob-1}. 

A huge number of researches on the single vacancy in Si crystal have
been done in the both experimental\cite{review1,prex1,prex2,review2} and
theoretical\cite{prth1,prth2,prth3,prth4} viewpoints. Early experiments
of the electron paramagnetic resonance (EPR)\cite{review1} revealed the
local lattice distortion around the positive charge state $V^+$ and
negative charge state $V^{-}$. Schl${\rm{\ddot u}}$ter {\it et al.}\cite{prth3,prth4} predicted that the negative-{\it U} effect due to
the Jahn-Teller distortion results in the $V^{++}$ groundstate with the
excited $V^{+}$ state; which is the well-known standard theory of the
single vacancy in Si crystal and is experimentally confirmed by Watkins
and Troxell\cite{prex1}. The negative charge state $V^{-}$ with the
Jahn-Teller distortion has also been confirmed by the EPR\cite{review1}
and the electron nuclear double resonance (ENDOR)\cite{prex2}
measurements. Recent first-principle
calculations\cite{fpc-1,fpc-2,fpc-3,fpc-4} with supercells up to
1000-atoms have shown that the calculated symmetry of the relaxed
vacancy state depends on computational details such as supercell size
and k-point sampling. Although the detailed symmetry of the charge state
is still controversial issue, the first-principle calculations have
concluded that the degeneracies of the charge states are resolved due to
the Jahn-Teller distortions except the doubly positive charge state
$V^{++}$\cite{fpc-4}. More recently, the elastic properties in a Si
vacancy have also been discussed by using the first-principle
calculations\cite{rfpc-1,rfpc-2}.

In the EPR experiments\cite{review1}, the vacancies were intentionally
created by electron-beam and $\gamma$-ray irradiations, and then, the
observed local distortions in the EPR are considered to be caused by the
cooperative Jahn-Teller effect among the highly irradiated vacancies
being strongly coupled each other\cite{ob-1}. Such vacancies are well
described by the first-principle calculations with supercells where the
vacancy concentration is $\sim 1/1000$. In the ultrasonic experiments,
however, the charge state of the thermally created vacancies with the
extremely low concentration in the FZ Si crystal, which is believed less
than $10^{15}$ cm$^{-3}$, shows no sign of the local distortion,
although a small inter-vacancy coupling $\sim$ 2 K is also
observed\cite{ob-1}. 

When only a single vacancy exists in the infinite Si crystal,
$T_d$-point symmetry should be preserved against the Jahn-Teller
distortions and the 3-fold orbital degeneracy should remain in the
vacancy groundstates. Even in the case with a finite vacancy
concentration, such a single vacancy state without the Jahn-Teller
distortions is expected to be a good starting point to describe the
properties for finite temperature, where the cooperative Jahn-Teller
effects due to the small inter-vacancy coupling are not
significant. Such situation is realized in the FZ Si crystal above at
least 20 mK and are observed in the ultrasonic experiments. Therefore we
need a theoretical study for a single vacancy in infinite Si crystal
without the Jahn-Teller distortions to describe the low temperature
elastic softening in the FZ Si crystal.

The similar elastic softenings have been widely observed in the 4$f$
electron systems\cite{rc-1,rc-2}, where the Curie-like behavior of the
quadrupole susceptibilities due to the orbital (quadrupole) degeneracy
of the 4$f$ electrons are responsible for the elastic softenings via the
quadrupole-strain coupling. Similarly to the 4$f$ electron systems, the
orbital degeneracy of the localized vacancy states in the Si crystal is
expected to result in the low temperature softenings observed in the
ultrasonic experiments. Recently, Matsuura and Miyake\cite{so} have
studied the quadrupole susceptibility due to a Si vacancy on the basis
of a cluster model for the dangling-bond orbitals in the Si vacancy
\cite{ob-1}. Yamakawa {\it et al.}\cite{rs-1,rs-2} have studied the
effect of the coupling between the electrons of the dangling-bonds and
the Jahn-Teller distortions by using the exact diagonalization for the
cluster model developed by Schl${\rm{\ddot u}}$ter {\it et
al.}\cite{prth3,prth4} and have revealed that the degeneracy of the
vacancy state remains against the Jahn-Teller effect because of the
strong quantum fluctuation due to the non-adiabatic effect, in contrast
to the case with the adiabatic approximation\cite{prth3,prth4}. However,
the effect of the spatial extension of the vacancy state, which is
important to determine the absolute value of the elastic softening, has
not been discussed in above cluster models. The experimental
group\cite{ob-1} has claimed that the widely extended vacancy state with
the effective radius \cite{prth4} $a\sim 5~{\rm\AA}$ is responsible for
the huge quadrupole susceptibility in proportion to $a^4$ which
contributes to the elastic constant via the quadrupole-strain coupling.

In this study, we focus on the $V^{0}$ charge state which has
non-magnetic groundstate and is realized in the non-doped Si as
confirmed by the ultrasonic experiment\cite{ob-1}. In this case, the
spin-orbit interaction, which is crucial for the magnetic groundstates
in the $V^{+}$ and $V^{-}$ states, is irrelevant. In the groundstate of
$V^{0}$, the two electrons are occupied in the 3-fold degenerate orbital
states with intra-orbital and inter-orbital spin-singlet configurations
which are responsible for the electric multipole susceptibilities such
as the quadrupole and octupole.

The purpose of this paper is to clarify the effect of the spatial
extension of the vacancy state on the quadrupole susceptibility which
contributes to the elastic constant at low temperature. 
For this purpose, we determine the electronic state around a single
vacancy in the infinite Si crystal by using the Green's function
approach\cite{G-1,G-2}. By virtue of this approach, we can discuss the
electronic state up to $30~{\rm \AA}$ from the vacancy site, where more
than 4000 Si atoms are included. Using the obtained Green's
functions, we calculate the quadrupole susceptibility on the basis of
the linear response theory. 

\section{Model and Formulation}
The model Hamiltonian consists of the tight-binding Hamiltonian $H_{0}$
and a vacancy potential $H_{\rm{v}}$ given by
\begin{eqnarray}
&&H=H_{0}+H_{\rm{v}}, \\
 &&H_{0} = \sum_{ij}\sum_{\alpha \beta} t_{ij}^{\alpha\beta}
 c^{\dagger}_{i\alpha}c_{j\beta}  = \sum_{{\bf k}}\sum_{m=1}^{8}
 \epsilon_{{\bf k} m}c^{\dagger}_{{\bf k} m}c_{{\bf k} m},
 \ \ \ \ \label{H0} \\ 
 &&H_{\rm{v}} = \Delta\sum_{\alpha} c_{0\alpha}^{\dagger}c_{0\alpha},
\end{eqnarray}
where $c_{i\alpha}^{\dagger}$ is a creation 
operator for an electron at site $i$ and orbital $\alpha~(={\it
s,p_x,p_y,p_z}$), and $c_{{\bf k} m}^{\dagger}$ is that for wave vector
${\bf k}$ and band $m~(=1\sim 8$). In eq. (\ref{H0}), 
the tight-binding parameters $t^{\alpha \beta}_{ij}$ are written by the Slater-Koster parameters and
determined so as to fit the tight-binding band energies $\epsilon_{{\bf k} m}$ 
to the LDA band energies\cite{LDA} for the Si crystal as shown in Fig. 1. 
The explicit values of the parameters are as follows: atomic levels
$\epsilon_{{\rm s}}=-4.778~\rm{eV},~\epsilon_{{\rm p}}=1.218~\rm{eV}$,
1st-neighbor hopping integrals ${\it t}_{{\rm
ss1}}=-2.104~\rm{eV},~{\it t}_{{\rm sp1}}=-1.788~\rm{eV},~{\it t}_{{\rm
pp\sigma1}}=-2.810~\rm{eV},~{\it t}_{{\rm pp\pi 1}}=-0.743~\rm{eV}$, 2nd-neighbor hopping integrals, which gives hopping between the dangling
bonds, $~{\it t}_{{\rm ss2}}=0.092~\rm{eV},~{\it t}_{{\rm
sp2}}=0.112~\rm{eV},~{\it t}_{{\rm pp\sigma 2}}=-0.386~\rm{eV},~{\it
t}_{{\rm pp\pi 2}}=-0.116~\rm{eV}$. 
The vacancy potential $H_{\rm{v}}$  excludes electrons from the
vacancy site by raising the energy levels $\Delta$ for the orbitals
belong to the vacancy site. For $\Delta\longrightarrow\infty$, no
electron exists at the vacancy site and then an effective vacancy state
is realized. 

\begin{figure}[htb]
\begin{center}
\includegraphics[height=6.0cm]{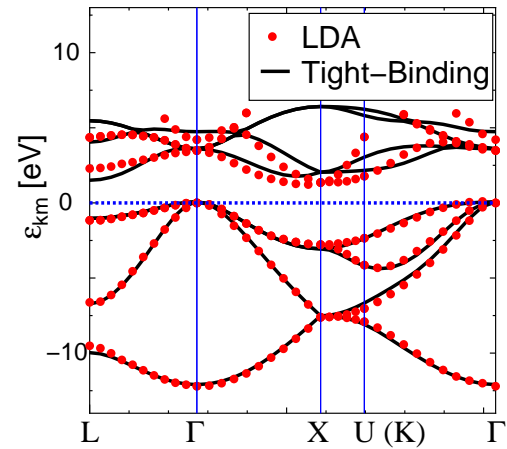}
\caption{(Color online) The band structure for the Si crystal calculated from the
 LDA \cite{LDA} (closed circles) and from the tight-binding model (solid
 lines).} 
\end{center}
\end{figure}

In the absence of the vacancy ($\Delta=0$), the Green's function for the
perfect crystal is described as
\begin{equation}
G^{0\alpha\beta}_{ij}(z)=\sum_{{\bf k}m}\frac{u_{\alpha m}({\bf k})u_{\beta m}^{*}({\bf k})}{z-\epsilon_{{\bf k}m}}e^{i{\bf k}\cdot\left({\bf r}_i-{\bf r}_j\right)} \label{G0}
\end{equation}
where $u_{\alpha m}({\bf k})$ is the eigenvector for the energy band $\epsilon_{{\bf k}m}$ and orbital $\alpha$ given in eq. (\ref{H0}). 
In the presence of the vacancy ($\Delta\ne 0$), the Green's function 
is obtained by
solving the Dyson's equations which can be written in the $4\times4$ matrix representation as
\begin{equation}
{\rm{\bf G}}_{ij}={\rm{\bf G}}^{0}_{ij}+{\rm{\bf G}}^{0}_{i0}
 {\bf \Delta}{\rm{\bf G}}_{0j}, \label{Dyson}
\end{equation}
with the vacancy potential matrix 
$({\bf\Delta})_{\alpha\beta}=\Delta\delta_{\alpha\beta}$, 
where $({\rm{\bf G}}^0_{ij})_{\alpha\beta}=G^{0\alpha\beta}_{ij}$ 
is the Green's function for $\Delta=0$ given in eq. (\ref{G0}) 
and $({\rm{\bf G}}_{ij})_{\alpha\beta}=G^{\alpha\beta}_{ij}$ is 
the corresponding Green's function for $\Delta\ne 0$. 
In the limit $\Delta\to \infty$, ${\rm{\bf G}}_{ij}\to 0$ with $i=0$ 
and/or $j=0$, and then
\begin{equation}
{\rm{\bf G}}^{0}_{0j}+{\rm{\bf G}}^{0}_{00}{\bf \Delta}
{\rm{\bf G}}_{0j} \rightarrow 0. \label{G00}
\end{equation}
By using eq. (\ref{G00}) in eq. (\ref{Dyson}), ${\rm{\bf G}}_{ij}$ is obtained 
by ${\rm{\bf G}}^{0}$ as
\begin{equation}
{\rm{\bf G}}_{ij}={\rm{\bf G}}^{0}_{ij}-{\rm{\bf G}}^{0}_{i0}\left({\rm{\bf G}}^{0}_{00}\right)^{-1}{\rm{\bf G}}^{0}_{0j}. \label{Gij}
\end{equation}
According to the Lehmann representation,
$G^{\alpha\beta}_{ij}$ is described by using the spectral function
$A_{i\alpha j\beta}^l$ and the excitation energy $E_{l}$, as 
\begin{equation}
G^{\alpha\beta}_{ij}(z)=\sum_{l}\frac{A_{i\alpha j\beta}^l}{z-E_l}.
\label{Lehmann}
\end{equation}

\section{Results}
\subsection{Density of States}
Now, we calculate $G^{0\alpha\beta}_{ij}$ in eq. (\ref{G0}) by 
performing the $\rm{\bf k}$ summation with $20\times 20\times 10 =4000$ 
mesh points, and substitute it into eq. (\ref{Gij}) to obtain 
$G^{\alpha\beta}_{ij}$ which yields $A_{i\alpha j\beta}^l$ 
and $E_l$ in eq. (\ref{Lehmann}). 
By using the obtained values of $A_{i\alpha j\beta}^l$ and $E_l$, 
we calculate the local density of states (DOS) at site $i$ in the
presence of the vacancy, $\rho_i(\omega)=\sum_{l\alpha}A_{i\alpha
i\alpha}^{l}\delta(\omega-E_l)$. 
Fig. 2 shows $\rho_i(\omega)$ together with the DOS for the perfect
crystal without vacancy, where the top of the valence band is set to the
energy origin. We find two remarkable localized levels : one sits in the
band gap and the other sits in the valence band. By summing the contribution
to each level from all sites, the total weight of each state is 3 for
the former level and 1 for the latter level. Therefore, we find that 
the former and 
the latter levels correspond to $T_2$ triplet states with the energy
$E_{T_2}=0.44$ eV and $A_1$ singlet state with the energy
$E_{A_1}=-0.12$ eV, respectively. These localized levels are occupied by
4 electrons in the $V^0$ state and by 3 electrons in
the $V^+$ state, respectively. In the both cases, the chemical
potential $\mu$ is close to the $T_{2}$ level at low temperature. Then in
the following, we focus on the $T_2$ triplet states which exclusively
contribute to thermodynamic quantities such as the quadrupole
susceptibilities at low temperature.

\begin{figure}[t]
\begin{center}
\vspace{0mm}
\includegraphics[height=13.0cm]{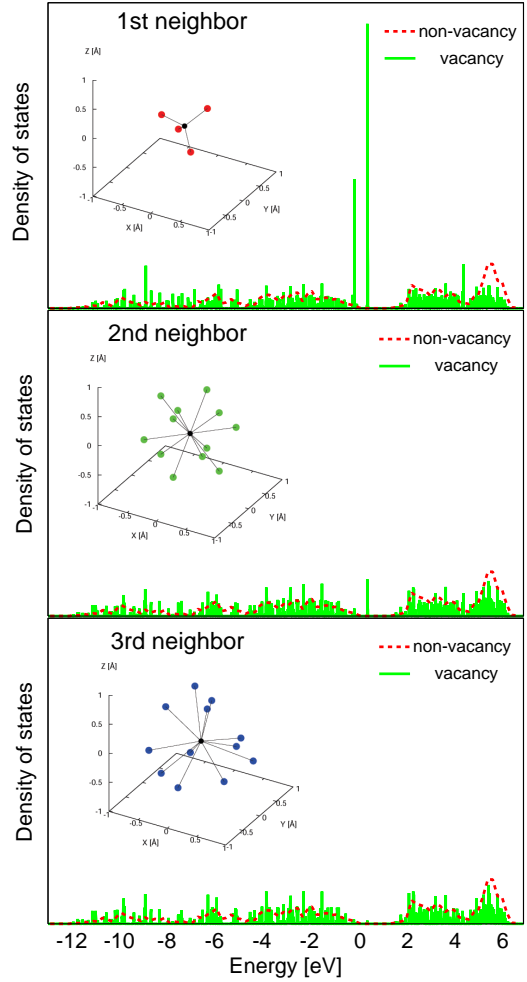}
\caption{(Color online) The local density of states in the presence of the vacancy
 (solid lines) together with those for the perfect crystal without
 vacancy (broken lines) at 1st (top), 2nd (middle) and 3rd (bottom)
 neighbor Si sites from the vacancy site.}
\end{center}
\end{figure}
\vspace{0mm}

The inset of Fig. 3 shows the local DOS for the $T_2$ triplet states at
site $i$, $N_{T_2}^i=\sum_{\alpha}A_{i\alpha i\alpha}^{T_2}$, as a function of the distance $R_i$ from the vacancy
site. We can see that, with increasing $R$, $N_{T_2}^i$ exponentially
decreases with decay length $\xi$ of several {\AA} accompanied by a
complicated oscillation. We also calculate the integrated local DOS for
$T_2$ triplet states up to the radius $R_c$ from the vacancy site,
$N_{T_2}=\sum_{i}^{R_c}N_{T_2}^i$, which is plotted as a function of
$R_c$ in Fig. 3. When $R_c\rightarrow\infty$, we can see
$N_{T_2}\rightarrow 3$ as expected. We note that a large part
($\sim80$ \verb|%|) of $T_2$ triplet states is concentrated on the region
$R< 5$ {\AA} with 34 atoms, but a small part ($\sim20$ \verb|%|) of
the widely extended vacancy states $R= 5\sim 20$ {\AA} with
$\sim10^3$ atoms play crucial roles for thermodynamic quantities such
as the quadrupole susceptibilities as will be shown later.

\begin{figure}[t]
\begin{center}
\vspace{0mm}
\includegraphics[height=6.2cm]{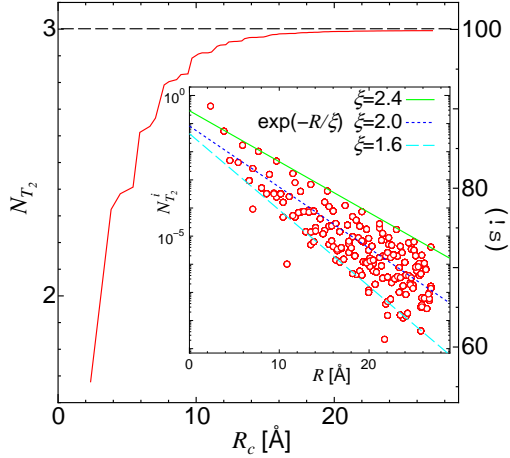}
\caption{(Color online) The integrated local DOS for $T_2$ triplet states up to the
 radius $R_c$ from the vacancy site. The inset shows the local DOS for
 the $T_2$ triplet states as a function of the distance $R$ from the
 vacancy site. The solid, dot and dashed lines are functions
 $\exp({-R}/{\xi})$ with $\xi=2.4$,  $\xi=2.0$ and $\xi=1.6$,
 respectively.}
\end{center}
\end{figure}

\subsection{Quadrupole susceptibility}
The low temperature elastic softening is considered to be caused by the
interaction between the electric quadrupole and the elastic
strain\cite{ob-1}, which is explicitly given by
\begin{align}
H_{QS}(t)&=-\sum_{\Gamma}g_{\Gamma}\tau_{\Gamma}\epsilon_{\Gamma}(t)\label{eq:H} 
\end{align}
with a quadrupole-strain coupling constant $g_{\Gamma}$. Here, $\epsilon_{\Gamma}(t)$ is a
strain with mode $\Gamma$ excited by the ultrasonic wave and couples to 
a quadrupole operator, 
\begin{equation}
\tau_{\Gamma}= e\sum_{ij}\sum_{\alpha \beta}\langle i\alpha |\phi_{\Gamma}({\bf r}) |j\beta \rangle
c_{i\alpha}^{\dagger}c_{j\beta},\label{eq:op} 
\end{equation}
with $\phi_{\Gamma}({\bf r})$, given by
\begin{eqnarray}
&\phi_{\Gamma}({\bf r})\equiv 
\left\{ \begin{array}{lcc}
(3z^2-r^2)/\sqrt{3}&{\rm for}&\tau_{\Gamma}=O_2^0 ~~(\epsilon_u) \\
x^2-y^2&{\rm for}&\tau_{\Gamma}=O_2^2 ~~(\epsilon_v),
\end{array} \right.\nonumber\\
&\phi_{\Gamma}({\bf r})\equiv
\left \{ \begin{array}{lcc}
yz&{\rm for}&\tau_{\Gamma}=O_{yz}~~(\epsilon_{yz})\\
zx&{\rm for}&\tau_{\Gamma}=O_{zx}~~(\epsilon_{zx})\\
xy&{\rm for}&\tau_{\Gamma}=O_{xy}~~(\epsilon_{xy}),
\end{array} \right.\nonumber
\end{eqnarray}
where the electric quadrupoles of $O_{2}^{0},O_{2}^{2}$ couple to the
tetragonal strain $\epsilon_u,\epsilon_v$ and $O_{yz},O_{zx},O_{xy}$
couple to the trigonal strain $\epsilon_{yz},\epsilon_{zx},\epsilon_{xy}$, 
respectively. Since the radius of the
atomic orbital of each Si atom at site $i$ is smaller than the distance
$R_i$ from the vacancy site to the Si atom, we can approximate $\langle
i\alpha |\phi_{\Gamma}({\bf r}) |j\beta \rangle\approx
\phi_{\Gamma}({\bf r}_i)\delta_{ij}\delta_{\alpha\beta}$, and then eq. (\ref{eq:H}) is rewritten as
\begin{align}
H_{QS}(t)&=-g_{\Gamma}e\epsilon_{\Gamma}(t)\sum_{i\alpha}\phi_{\Gamma}({\bf r}_i) c_{i\alpha}^{\dagger}c_{i\alpha}.
\end{align}

The relationship between the elastic constant $C_{\Gamma}(T)$ and the
quadrupole susceptibility $\chi_{\Gamma}(T)$ at temperature $T$ is given
in the second order perturbation w.r.t. $g_{\Gamma}$: 
\begin{align}
C_{\Gamma}(T)=C_{\Gamma}^0-N_{\rm v}g_{\Gamma}^2\chi_{\Gamma}(T)\label{eq:C},
\end{align}
where $C_{\Gamma}^0$ is the background of the elastic constant and
$N_{\rm v}$ is the number of vacancies. 
We note that the quadrupole susceptibilities of $O_{2}^{0},O_{2}^{2}$
contribute to the elastic constant of $(C_{11}-C_{12})/2$, while those
of $O_{yz},O_{zx},O_{xy}$ contribute to $C_{44}$.
In the linear response theory,
the quadrupole susceptibility is given by
\begin{align}
&\chi_{\Gamma}(T)={\rm Re}\int_{0}^{\infty} dt ~e^{-0_{+}t}\langle [\tau_{\Gamma}^{\dagger}(t),\tau_{\Gamma}]\rangle  \nonumber\\
&=-e^2\sum_{ll'}\sum_{ij\alpha\beta}\phi_{\Gamma}({\bf r}_i)\phi_{\Gamma}({\bf r}_j)A^{l}_{i\alpha j\beta}A^{l'}_{j\beta i\alpha}\frac{f(E_l)-f(E_{l'})}{E_{l}-E_{l'}}\nonumber
\end{align}
with the fermi distribution function 
$f(E)=1/(e^{\beta(E-\mu)}+1)$, where $\chi_{\Gamma}(T)$
consists of the Curie term ($E_{l}=E_{l'}$) and the Van-Vleck term
($E_{l}\neq E_{l'}$). As the Van-Vleck term is almost $T$-independent at
low temperature, we focus only on the Curie term due to the degenerate
$T_2$ triplet states ($E_l=E_{l'}=E_{T_2}$), which is explicitly given by
\begin{align}
\chi_{\Gamma}(T)&=e^2\sum_{ij\alpha\beta}\phi_{\Gamma}({\bf
 r}_i)\phi_{\Gamma}({\bf r}_j)A^{T_{2}}_{i\alpha j\beta}A^{T_2}_{j\beta
 i\alpha}\frac{f(E_{T_2})f(-E_{T_2})}{T}\nonumber\\
&=\frac{K_{\Gamma}}{T}F(n)  \label{eq:K1} 
\end{align}
with $F(n)=n(6-n)/36$, 
where $n=6f(E_{T_2})$ is the occupation number of electrons in the $T_2$
triplet states, and $n=2$ for the $V^0$ state. Substituting eq. (\ref{eq:K1}) into eq. (\ref{eq:C}), we obtain
the elastic constant which shows the reciprocal $T$ dependence,
$C_{\Gamma}(T)\propto -N_{{\rm v}}g_{\Gamma}^2K_{\Gamma}F(n)/T$, where
the absolute value of the elastic softening is determined by the Curie
constant of the quadrupole susceptibility given by
\begin{align}
K_{\Gamma}=e^2\sum_{ij}\sum_{\alpha\beta}\phi_{\Gamma}({\bf r}_i)\phi_{\Gamma}({\bf r}_j)A^{T_2}_{i\alpha j\beta}A^{T_2}_{j\beta i\alpha}.\label{eq:K}
\end{align}

\begin{figure}[t]
\begin{center}
\vspace{0mm}
\includegraphics[height=6.2cm]{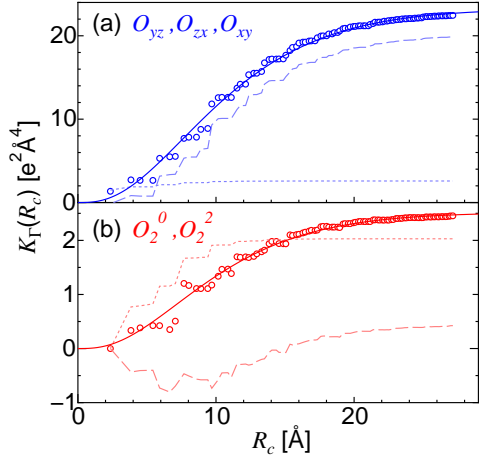}
\caption{(Color online) The Curie constant $K_{\Gamma}$ for the trigonal mode
 ($O_{yz},O_{zx},O_{xy}$) (a) and the tetragonal mode
 ($O_{2}^{0},O_{2}^{2}$) (b) as functions of $R_c$. The open circles, 
 dotted and dashed lines  correspond to the total, site
 diagonal and site off-diagonal  contributions, respectively. The solid
 lines show fitting functions for the total contributions (see in the text). 
}
\end{center}
\end{figure}
\vspace{0mm}

By using eq. (\ref{eq:K}), we calculate $K_{\Gamma}$ with the site
summation ($i,j$) up to the distance $R_c$ from the vacancy site as done
for the calculation of $N_{T_2}$, where $K_{\Gamma}$ consists of the
site diagonal contribution ($i=j$) and the site off-diagonal
contribution ($i\neq j$). Fig. 4 shows $K_{\Gamma}$ together with the
site diagonal and site off-diagonal contributions for $K_{\Gamma}$ as
functions of $R_c$. To estimate the $R_c\rightarrow\infty$ extrapolated
value of $K_{\Gamma}$, we assume a fitting function
$K_{\Gamma}(R_c)\approx
K_{\Gamma}(\infty)[1-\{R_c^2/(2\xi_{_{\Gamma}}^2)+R_c/\xi_{_{\Gamma}}+1\}e^{-R_c/\xi_{_{\Gamma}}}]$\cite{fit}.
By comparing the calculated value of $K_{\Gamma}$ with the fitting
function as shown in Fig. 4, we obtain the  extrapolated values: 
$K_{\Gamma}(\infty)=23.28~e^2${\AA}$^4$ (trigonal) and 
$K_{\Gamma}(\infty)=2.51~e^2${\AA}$^4$ (tetragonal), 
together with the decay length: $\xi_{_{\Gamma}}=3.77$  (trigonal) and 
$\xi_{_{\Gamma}}=3.56$ {\AA} (tetragonal). We find that $K_{\Gamma}$
for the trigonal mode is about 10 times larger than that for the
tetragonal mode as shown in Fig. 4, where the site diagonal contribution
for the trigonal mode is almost the same as that for the tetragonal
mode, while the site off-diagonal contribution for the trigonal mode is
much larger than that for the tetragonal mode. The result is consistent 
with the experimental result, where the softening of $C_{44}$ is considerably 
larger than that of $(C_{11}-C_{12})/2 $\cite{ob-1}, if the 
coupling constants $g_{\Gamma}$ for the both modes, which have not been 
determined so far,  are of the same order of magnitude. 

Remarkably, the widely extended vacancy states on the region $R= 5 \sim
20$ {\AA} contribute to about 90 \verb|%| of the absolute value of
$K_{\Gamma}$ as shown in Fig. 4, although they contribute to only 20
\verb|%| of the local DOS as shown in Fig. 3. In fact, the decay length
$\xi_{_{\Gamma}}$ for the Curie constant $K_{\Gamma}$ is about twice
larger than  the decay length $\xi$ for the local DOS (see also the inset 
of Fig. 3). 
We note that the effective 
quadrupole moments estimated by $(K_{\Gamma})^{1/2}$ are 4.82
$e${\AA}$^2$ (trigonal) and 1.58 $e${\AA}$^2$ (tetragonal), which are
considerably larger than molecular quadrupole moments of the order of
0.1 $e${\AA}$^2$ \cite{quad}.

\subsection{Multipole susceptibility}
It is expected that the widely extended vacancy states also yield the
extreme enhancement of the other multipole susceptibilities. Then, we
also calculate the electric dipole and octupole susceptibilities by
using the same method as the quadrupole susceptibilities. The Curie
constants $K_{\Gamma}$ for the corresponding multipole susceptibilities
are obtained by replacing $\phi_{\Gamma}({\bf r})$ in eq. (\ref{eq:op})
with
\begin{eqnarray}
\phi_{\Gamma}({\bf r})\equiv
\left\{ \begin{array}{lll}
x&\hspace{0mm}{\rm for}&\hspace{0mm}\tau_x \\
y&\hspace{0mm}{\rm for}&\hspace{0mm}\tau_y \\
z&\hspace{0mm}{\rm for}&\hspace{0mm}\tau_z
\end{array} \right. \nonumber
\end{eqnarray}
for the electric dipole and
\begin{eqnarray}
\phi_{\Gamma}({\bf r})\equiv\left\{ \begin{array}{lll}
xyz&\hspace{0mm}{\rm for}&\hspace{0mm}\tau_{xyz}  \\
x(3x^2-r^2)/\sqrt{3}&\hspace{0mm}{\rm for}&\hspace{0mm}\tau_{x}^{\alpha} \\
y(3y^2-r^2)/\sqrt{3}&\hspace{0mm}{\rm for}&\hspace{0mm}\tau_{y}^{\alpha} \\
z(3z^2-r^2)/\sqrt{3}&\hspace{0mm}{\rm for}&\hspace{0mm}\tau_{z}^{\alpha} \\
x(y^2-z^2)&\hspace{0mm}{\rm for}&\hspace{0mm}\tau_{x}^{\beta} \\
y(z^2-x^2)&\hspace{0mm}{\rm for}&\hspace{0mm}\tau_{y}^{\beta} \\
z(x^2-y^2)&\hspace{0mm}{\rm for}&\hspace{0mm}\tau_{z}^{\beta}
\end{array} \right.\nonumber
\end{eqnarray}
for the electric octupole.

Fig. 5 shows $K_{\Gamma}$ for several multipole susceptibilities as
functions of $R_c$. It is found that $K_{\Gamma}$ for one order higher
multipole susceptibilities show about 10 times larger enhancement. In
the octupole susceptibilities, the site diagonal contributions for
$\tau_{xyz},\tau^{\alpha},\tau^{\beta}$ are the same order of 10
$e^2${\AA}$^6$ (not shown), while the site off-diagonal contribution
for $\tau^{\alpha}$ is much larger than that for the other mode resulting
in the extreme enhancement of $\tau^{\alpha}$. We note that $K_\Gamma$ for
$\tau^{\beta}$ decreases with increasing $R_c$ because its site
off-diagonal contribution is negative.

\begin{figure}[t]
\begin{center}
\vspace{0mm}
\includegraphics[height=6.2cm]{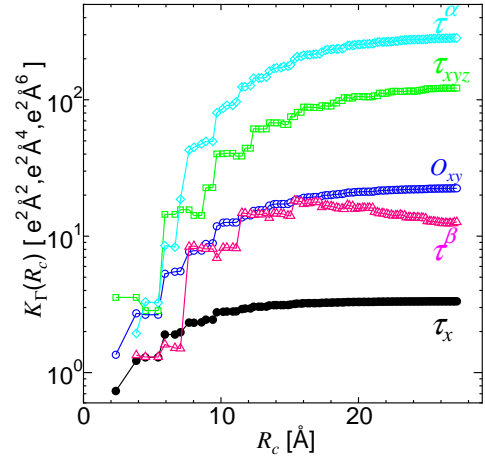}
\caption{(Color online) The Curie constants $K_{\Gamma}$ for various multipole
 susceptibilities as functions of $R_c$. The closed circles ($\bullet$),
 open circles ($\circ$), squares (${\scriptscriptstyle \square}$),
 diamonds ($\diamond$) and triangles (${\scriptstyle \triangle}$)
 represent the $K_{\Gamma}$ for the dipole $\tau_x$, quadrupole $O_{xy}$,
 octupole $\tau_{xyz},\tau^{\alpha}$ and $\tau^{\beta}$ susceptibilities,
 respectively.}
\end{center}
\end{figure}
\vspace{0mm}

\section{Summary and Discussion}
In summary, we have investigated the electronic state around a single
vacancy in infinite Si crystal on the basis of the Green's function
approach. It has been found that the $T_2$ triplet vacancy states within
the band gap are widely extended up to 20 {\AA} and are
responsible for the extreme enhancement of the Curie constant of the
quadrupole susceptibilities resulting in the elastic softening at low
temperature. The Curie constant for the trigonal mode
($O_{yz},O_{zx},O_{xy}$) is considerably larger than that for the
tetragonal mode ($O_{2}^{0},O_{2}^{2}$). These results are consistent
with the low temperature elastic softening observed in the ultrasonic
experiments.\cite{ob-1} We have also calculated the other multipole
susceptibilities and found that the remarkable enhancement of the Curie
constant for the octupole susceptibilities especially in $\tau^{\alpha}$
mode; which is expected to be observed in future experiments.

In the present study, we have assumed that the $T_d$-point symmetry
remaining the orbital degeneracy is preserved in a single vacancy in
infinite Si crystal. Even in this case, a lattice relaxation with
keeping the symmetry might take place; which results in a modification
of the tight-binding parameters around the vacancy. To discuss this
effect, we need a first-principle calculation with keeping the
symmetry. A recent molecular dynamics simulation combined with a
first-principle calculation revealed that such a high symmetric vacancy
state is realized at a finite temperature\cite{rfpc-1}. The explicit
calculation of the quadrupole susceptibility including the effect of the
lattice relaxation is an important future
problem. In addition, the explicit estimation of the quadrupole-strain
coupling $g_\Gamma$ is also important to determine the absolute value of
the elastic softening.

At a finite vacancy concentration, the effect of the inter-vacancy
interaction is considered to be also important. In fact, the ultrasonic
experiments revealed that there exists the antiferro-type inter-vacancy
interaction of order of 2 K\cite{ob-1}. On the basis of the Green's
function approach employed in the present study, we can discuss the
effect of the inter-vacancy interaction by introducing the spatially
separated two vacancies in infinite Si crystal. Such a calculation is
under the way.

The effects of the Coulomb interaction and the coupling between
electrons and Jahn-Teller phonons, which have not been considered 
in the present study, are 
crucial to determine the many-body groundstate in a Si vacancy and have
been intensively discussed from the cluster model
calculations\cite{rs-1,rs-2}. 
Actually, the non-magnetic groundstate in $V^{0}$ with 3-fold orbital 
degeneracy has been obtained due to the effect of nonadiabatic couplings 
between electrons and Jahn-Teller phonons \cite{rs-1,rs-2}, 
in contrast to the present study where the spin degeneracy is 
unresolved as an artifact of the noninteracting calculation. 
In addition, the effect of the spin-orbit interaction neglected in this
study is also crucial for the magnetic groundstates of $V^+$ and $V^-$
which are expected to be realized in the B-doped\cite{so}
and P-doped Si, respectively. On the basis of the Green's functions 
for a Si vacancy obtained
by the present study, we can discuss the many-body effect by including
the selfenergy corrections due to the Coulomb interaction and the
electron-phonon coupling. The detailed results of the selfenergy
corrections in the presence of the spin-orbit interaction will be
reported in a subsequence paper.

\section*{Acknowledgments}
The authors thank T. Goto, H. Kaneta, Y. Nemoto, K. Mitsumoto, K. Miyake
and H. Matsuura for many useful comments and discussions. This work was
partially supported by the Grant-in-Aid for Scientific Research  from
the Ministry of Education, Culture, Sports, Science and Technology.

\end{document}